\documentclass[12pt]{iopart}
\usepackage{ifthen}
\newcommand{\usepdfs}{false}
\ifthenelse{\equal{\usepdfs}{true}}{
  \usepackage[pdftex]{graphicx}
}
{
  \usepackage[dvips]{graphicx}
}
\ifthenelse{\equal{\usepdfs}{true}}{
  
}
{
  
}

\usepackage{subfigure}
\usepackage{amssymb}

\newcommand{\ket}[1]{| #1 \rangle}
\newcommand{\bra}[1]{\langle #1 |}

\renewcommand{\u}{\uparrow}
\renewcommand{\d}{\downarrow}

\newcommand{\be}{\begin{equation}}
\newcommand{\ee}{\end{equation}}

\begin{document}

\title{Effective Hamiltonians for some highly frustrated magnets}

\author{Doron L. Bergman$^1$, Ryuichi Shindou$^1$, Gregory
  A. Fiete$^2$, and Leon Balents$^1$}
\address{${}^1$Department of Physics, University of California,
  Santa Barbara, CA 
93106-9530\\$^2$Kavli Institute for Theoretical Physics, University of 
California, Santa Barbara, CA 93106-4030}
%\ead{doronber@physics.ucsb.edu}

\begin{abstract}
%We analyze the easy axis degenerate perturbation theory for a family of frustrated XXZ models.
  In prior work, the authors developed a method of degenerate
  perturbation theory about the Ising limit to derive an effective
  Hamiltonian describing quantum fluctuations in a half-polarized
  magnetization plateau on the pyrochlore lattice.  Here, we extend this
  formulation to an arbitrary lattice of corner sharing simplexes of $q$
  sites, at a fraction $(q-2k)/q$ of the saturation magnetization, with
  $0<k<q$.  We present explicit effective Hamiltonians for the examples
  of the checkerboard, kagome, and pyrochlore lattices.  The consequent
  ground states in these cases for $k=1$ are also discussed.
\end{abstract}
%\date{\today}
%\pacs{75.10.-b,75.10.Jm,75.25.+z}

%75.10.Jm Quantized spin models
%75.10.-b General theory and models of magnetic ordering (see also 05.50.+q Lattice theory and statistics)
%75.25.+z Spin arrangements in magnetically ordered materials (including neutron and spin-polarized electron studies, synchrotron-source x-ray scattering, etc.) (for devices exploiting spin polarized transport, see 85.75.-d)
%75.60.-d Domain effects, magnetization curves, and hysteresis

%\email{doronber@physics.ucsb.edu}

\section{Introduction}
\label{sec:introduction}

Despite decades of theoretical and experimental work, frustrated
quantum magnets continue to be an exciting subject for current
research both experimentally and theoretically.  Analytic approaches
to such spin systems have, however, not often directly confronted
experiment.  Theoretically, the semi-classical spin-wave theory
(i.e $1/s$ expansion) has probably been most successful.  For
instance, quantum corrections to the classical
staggered magnetization are known to be small even for $s=1/2$ 
in the unfrustrated square lattice.
The utility of the $1/s$ expansion, however, diminishes rapidly as
frustration is increased.  

This situation is at its extreme in the class of maximally
geometrically frustrated structures, which includes the
two-dimensional kagome and checkerboard lattices, and the
three-dimensional pyrochlore lattice.  These structures have the
common feature that they can be decomposed into distinct ``corner
sharing'' {\sl simplexes}, clusters of spins in which all pairs are
connected by nearest neighbor bonds, such that the entire lattice is
covered by these simplexes, and different simplexes share at most one
site and no bonds.  The nearest-neighbor antiferromagnet on this
lattice has the property that its Hamiltonian can be written entirely
in terms of the sum of spins on each simplex.  At the classical level,
this implies a large degeneracy, since any change in configuration
which keeps this sum constant on each simplex cannot change the
energy.  This leads to considerable technical difficulties in the
$1/s$ expansion, which have been addressed in tour-de-force work by
Henley and collaborators~\cite{Hizi:prb05}.  An unfortunate outcome of this work is that
the leading order ($O(1/s^0)$) corrections (spin-wave
zero point energy) do not fully split the classical degeneracy in most
interesting cases.  Given the challenges already present at
$O(1/s^0)$, it is not surprising that higher order corrections in
$1/s$, which would be necessary to resolve the degeneracy fully and
determine the quantum ground state for large $s$, are so far not
available.

In recent work, we have demonstrated that the resolution of the
classical degeneracy {\sl can} be understood in an alternate approach,
based on perturbation theory about the limit of strong Ising exchange
anisotropy.  That is, we consider the XXZ model 
\begin{equation}
  \label{eq:xxz}
  {\mathcal H} = J \sum_{\langle i j \rangle} S^z_i S^z_j 
+ \frac{J}{2} \alpha \sum_{\langle i j \rangle} \left( S_i^+ S_j^- +
  h.c. \right)  - H \sum_j S^z_j \; ,
\end{equation}
and carry out perturbation theory in $\alpha$ for the degenerate
ground state manifold.  This is expected to be a reasonable
approximation in many problems of interest.  First, the semiclassical
approach demonstrates rather generally that in these models quantum
fluctuations favor {\sl collinear} ordered states, in which all spin
expectation values are aligned along a particular, e.g. $S^z$ axis.
Choosing Ising anisotropy only selects this axis, but does not
prejudice the ordering beyond this choice.  Second, some of the most
interesting applications are to {\sl magnetization plateaus}, which
often appear in frustrated magnets (e.g in pyrochlores~\cite{Ueda:prb06}).  General arguments (see e.g.
\cite{Bergman:prl05}) imply that ordering is again collinear on such plateaus.
Moreover, since such a plateau occurs in a substantial applied field,
the component of each spin parallel to this field is clearly on
average larger than its transverse ones.  Furthermore, the
introduction of Ising anisotropy does not modify the symmetry of the
Hamiltonian in this case.  In \cite{Bergman4}, we have
demonstrated explicitly how to carry out easy-axis degenerate
perturbation theory for a magnetization plateau on the pyrochlore
lattice at half the saturation polarization.  The result was an
effective quantum ``dimer'' Hamiltonian describing the splitting of
the degenerate manifold of plateau states.

In this paper, we describe the generalization of these results to the
lattices described above, for various zero field and plateau states.
Our results are obtained under the assumption that each corner-sharing
simplex contains $q$ sites, and that the Ising ground state manifold is
comprised of states with $k$ ``minority'' ($S^z=-s$) and $q-k$
``majority'' ($S^z=+s$) spins per simplex.  The cases mentioned above
correspond to $q=3,k=1$ (kagome lattice at magnetization $M=1/3 M_s$),
$q=4,k=2$ (checkerboard and pyrochlore lattices at $M=0$), $q=4, k=1$
(checkerboard and pyrochlore lattices at $M=1/2 M_s$).  For the $k=1$
cases, the corresponding effective Hamiltonians are generalized
``quantum dimer models'', and we discuss their ground states based on
known results and simple arguments.  The solution for the ground states
of the $k=2$ models (corresponding to zero magnetization) on the
pyrochlore and checkerboard lattices are left for future work.

The remainder of the paper is organized as follows.  In
section~\ref{sec:Model}, we describe the derivation of the effective
Hamiltonians.  In section~\ref{sec:results}, we give explicit forms for
the above-mentioned lattices. In section~\ref{sec:lowest-energy-states}
we discuss the ground states of the effective models in the $k=1$
cases.

\section{Resume of Method}
\label{sec:Model}

In this section, we review the degenerate perturbation theory (DPT)
methods developed in \cite{Bergman4}.  Because the fine details
are given already in \cite{Bergman4}, we will highlight only the
main points and those modifications needed for the more general
applications in this paper.  

For any (apart from a set of measure zero) fixed field $H$ in the
Ising limit $\alpha=0$, the ground states of equation~\ref{eq:xxz}
comprise a massively degenerate manifold of configurations of constant
magnetization -- ``plateau states''.  Specifically, in these
configurations (as indicated in the introduction) all spins are
aligned along the field axis, i.e $S_i^z=s\sigma_i$, with
$\sigma_i=\pm 1$, and every simplex contains $k$ minority
($\sigma_i=-1$) spins and $q-k$ majority ($\sigma_i=+1$) spins.  We
seek an effective Hamiltonian in this degenerate subspace.

We follow the Brillouin-Wigner formulation of DPT.  The ground state
wavefunction satisfies the exact equation
\begin{equation}
  \label{eq:effshrod}
   \left[ E_0 + 
{\mathcal P} {\mathcal H}_1 \sum_{n=0}^{\infty} {\mathcal G}^n 
{\mathcal P} \right] |\Psi_0\rangle = E |\Psi_0\rangle = {\mathcal H}_{\textrm{eff}} |\Psi_0\rangle,
\end{equation}
where the operator ${\mathcal G} = \frac{1}{E - {\mathcal H}_0} \left(
  1 - {\mathcal P} \right) {\mathcal H}_1 $.  Here ${\mathcal
  H}_0=\left.{\mathcal H}\right|_{\alpha=0}$, and ${\mathcal
  H}_1={\mathcal H}-{\mathcal H}_0$.  Because the resolvent contains
the exact energy $E$, equation~(\ref{eq:effshrod}) is actually a non-linear
eigenvalue problem.  However, to any given order of DPT, $E$ may be
expanded in a series in $\alpha$ to obtain an equation with a true
Hamiltonian form within the degenerate manifold.  Each factor of
${\mathcal G}$ is at least of $O(\alpha)$ due to the explicit factor
in ${\mathcal H}_1$, with higher order corrections coming from the
expansion of $E$.  Following the strategy of \cite{Bergman4}, we
will obtain the lowest order non-constant diagonal and off-diagonal
terms in ${\mathcal H}_{\textrm{eff}}$.  Because we seek only the
lowest order term of each type, it is admissible to replace $E$ by
$E_0$ in ${\mathcal G}$, a considerable simplification.

With this replacement, we need only calculate the successive
applications of ${\mathcal G}$ and ${\mathcal H}_1$ in
equation~\ref{eq:effshrod} starting from some arbitary initial state.
This is facilitated by the simple geometry of the lattice, and by
symmetry.  The Hamiltonian in equation~\ref{eq:xxz} has a global $U(1)$
symmetry, corresponding to rotations about the z-axis. As such, the
z-component of the total magnetization $\sum_j S^z_j$ is a conserved
quantity at every stage of DPT.  This leads to some important
properties of the resolvent operator ${\mathcal R}=({\mathcal
  H}_0-E_0)^{-1}$:
\begin{equation}
  \label{eq:resolve}
  {\mathcal R}^{-1} = 
 \frac{J}{2} \sum_{i j} \Gamma_{i j} S_i^z S_j^z - H \sum_j S_j^z - E_0
 \; ,
\end{equation}
where $\Gamma_{ij}$ is the adjacency matrix, i.e. $\Gamma_{ij}=1$ if
$i,j$ are nearest-neighbors and $\Gamma_{ij}=0$ otherwise.  We will
need the action of ${\mathcal R}$ not on states in the ground state
manifold, but on an arbitrary virtual state reached in a DPT process.
We describe these states by integers $m_j$, with $0\leq m_j\leq 2s$,
such that $S_j^z=\sigma_j(s-m_j)$.  Conservation of magnetization
enforces $\sum_j m_j \sigma_j=0$ for all virtual states.  One may
readily confirm the additional identity 
\be
\sum_i \Gamma_{i j} \sigma_i = 2(q-2k) - 2 \sigma_j,
\label{eq:id1}
\ee
which results from the constraints on the spin configurations
contained on the two simplexes sharing site $j$.  Using these two
relations, the inverse resolvent can be rewritten as
\be\label{resolvent3}
{\mathcal R}^{-1} = 
\frac{J}{2} \sum_{i j} \Gamma_{ij} m_i m_j \sigma_i \sigma_j + 2 s J
\sum_j m_j 
\; .
\ee
Note that the resolvent only involves variables on the sites that are
modified in the DPT process, since all other $m_j = 0$.  With this
observation, we are ready to describe the calculations.

We first consider the off-diagonal term, which must cause transitions
from one plateau state to another.  To maintain the constraint
requires flipping spins $\pm s \rightarrow \mp s$ in a closed loop of
alternating $\pm s$ spins.  This is because an open string of spin
transfer changes the number of minority sites on the simplexes at the
two ends of the string: one suffers a deficit, and the other suffers a
surplus. Since ${\mathcal H}_1$ is a sum of spin transfer operators on
links of the lattice, one may readily deduce the order at which the
first off-diagonal term appears, without calculating it explicitly.
The smallest even-length non-trivial loop in the lattice
(non-retracing walk residing on more than one simplex) is of some even
length $L$ ($L = 6$ for pyrochlore and kagome, $L=4$ for Checkerboard)
and surrounds a plaquette of the lattice.  This loop contains $L/2$
minority sites that will be converted to majority sites. Each flipping
takes $2s$ operations of ${\mathcal H}_1$, so that the
off-diagonal term is of order $\alpha^{s L}$.  For large $s$, this
becomes of very high order and entirely negligible.  It can however be
important for moderate values of $s$.  

As is clear from this discussion, the lowest-order off-diagonal term
generates a single process: flipping all spins along one length-$L$
loop.  Thus it has the general form
\be {\mathcal H}_{\textrm{off diagonal}}
= - K_L(s) \alpha^{L s} J \sum_{\mathcal P} \left(
  \ket{\d\u\ldots\d\u} \bra{\u\d\ldots\u\d} + h.c. \right) \; .  \ee
What remains to be calculated is the coefficient $K_L(s)$ multiplying
the plaquette flipping operator.  In \cite{Bergman4}, an
iterative method was developed to calculate this amplitude by
considering successive applications of ${\mathcal H}_1$ on any initial
state.  Because all the spin-flips occurring in any given process are
on a single length-$L$ loop, the calculation is insensitive to the
global structure of the lattice, and depends only upon the length of
the plaquette.  This implies that, for hexagonal plaquettes, the
off-diagonal coeffcient is \emph{identical} to that calculated in
\cite{Bergman4}, even on the kagome lattice!  The values for
$K_6(s)$ for $s = 1/2,1,3/2,2,5/2$ were calculated in
\cite{Bergman4}.

For the square plaquette lattices, the calculation is actually
significantly simpler.  At a given stage in the DPT process, the
resolvent can be calculated using the number of times every
(alternating) link has undergone spin transfer up to that step.  For
the square plaquette, there are only two such links, as opposed to
three for the hexagonal plaquette.  After $\ell$ link operations, one
finds that the resolvent is given simply by
\be {\mathcal R}_{\ell}^{-1} = J \ell
(4s-\ell).
\ee
Applying this formula, the off-diagonal coefficient
is determined explicitly for any value of $s$:
\be K_4(s) =
\frac{4 s}{2^{4 s}} \frac{[(2 s)!]^2}{(4s-1)!}  .
\ee 

We now turn to the diagonal terms in DPT.  Unlike the off-diagonal
terms above, these occur at a fixed order in $\alpha$ independent of
$s$.  To understand at what order this occurs, first note that a
diagonal DPT process of order $n$, which starts out from some plateau
state, and returns to it in the end can have at most $n$ sites
modified, since each site that is modified must be modified again at
least once in order to return to its original configuration.  From the
structure of the resolvent discussed above, we say that at $n^{\rm
  th}$ order, the diagonal effective Hamiltonian is a sum of terms,
each of which involves at most $n$ Ising spin variables.  The
diagonal effective Hamiltonian therefore has the general structure
\be\label{Ising_func}
{\mathcal H}_{\rm eff}[\{\sigma_i\}]  = 
\sum_n  \sum_{G_n} \sum_{a_1 \ldots a_n}\!\!\! \left(\prod_{(ij)\in G_n} \Gamma_{a_i a_j}\right)
  f_{G_n}(\sigma_{a_1},\ldots,\sigma_{a_n}) ,
\ee
where $G_n$ denotes a ``graph'' of lines connecting the labels $a_1
\ldots a_n$ visualized as points, i.e. a set of unordered pairs of
these labels.  Note that the dependence upon the lattice geometry
enters only through $\Gamma_{ij}$.  This function can be derived from
direct evaluation of equation~\ref{eq:effshrod} using the simplified
resolvent, as outlined in \cite{Bergman4}.

Remarkably, for any lattice composed of corner-sharing simplexes, all
terms in equation~\ref{Ising_func} with $n<L$ are constants in the
ground-state manifold.  This was demonstrated in \cite{Bergman4}
for the case of the pyrochlore lattice (where $L=6$), but the
arguments follow more generally.  We outline the basic ideas,
referring to \cite{Bergman4} for details.  First, one can show
that all ``contractible'' diagrams, in which at least one point in
$G_n$ is connected to less than 2 other points, can be reduced to a
term of lower order.  This follows by explicitly summing over the
corresponding index (e.g. $a_n$), using the identity in
equation~\ref{eq:id1} and the constant total magnetization.  Second, one
shows that ``non-contractible'' diagrams with $n<L$ are constant by a
different argument.  First, one notes that the sites $a_i$ for which
$\prod_{(ij)\in G_n} \Gamma_{a_i a_j} \neq 0$ must form a compact
cluster, and for $n<L$ this cluster cannot span the smallest
non-trivial loop.  Summing over {\sl all} clusters satisfying
$\prod_{(ij)\in G_n} \Gamma_{a_i a_j} \neq 0$, one again obtains
constant.  This is essentially because any permutation of the cluster
sites $a_i$ gives another cluster, and the sum over all such
permutations can be reduced using the fact that the allowed spin
configurations of a simplex are all permutations of one reference
configuration (with $k$ minority spins on $q$ sites).  

The lowest non-constant term in the diagonal effective Hamiltonian
therefore comes at order $n=L$.  Indeed, by application of the same
arguments described above, only one distinct term at this order
is non-constant: the one for which $G_n$ consists of a single loop
connecting all labels, i.e. $\prod_{(ij)\in G_L} \Gamma_{a_i a_j}=
\Gamma_{a_1 a_2}\Gamma_{a_2 a_3}\cdots \Gamma_{a_{L-1}a_L}\Gamma_{a_L
  a_1}$.  This term can be explicitly evaluated, and has non-constant
contributions only when $(a_1,\cdots,a_L)$ lie sequentially along one
of the length $L$ loops of the physical lattice.  Evaluating $f_{G_n}$
at the spins of these sites leads to the final expression for the
diagonal energy function.

\section{Results}
\label{sec:results}

The diagonal effective Hamiltonian obtained above can in all cases be
written as a sum over plaquettes 
\be {\mathcal H}_{\textrm{diag}} =
\sum_{\mathcal P} {\mathcal E}_{\mathcal P}(\sigma_{{\mathcal
    P}1},\cdots,\sigma_{{\mathcal P}L}) \; ,  
\ee
where $\sigma_{{\mathcal P}1}\cdots\sigma_{{\mathcal P}L}$ are the $L$
sites ordered sequentially around the plaquette ${\mathcal P}$.  We
choose the function ${\mathcal E}$ to have cyclic symmetry.  For the
kagome and pyrochlore lattices, the plaquettes are hexagonal, bounded
by 6 links.  For the Checkerboard lattice, the plaquettes are square,
bounded by 4 links.

\begin{figure}
	\centering
	\subfigure[hexagonal plaquette]{
	\label{fig:plaquette_types2}
        \includegraphics[width=3.0in]{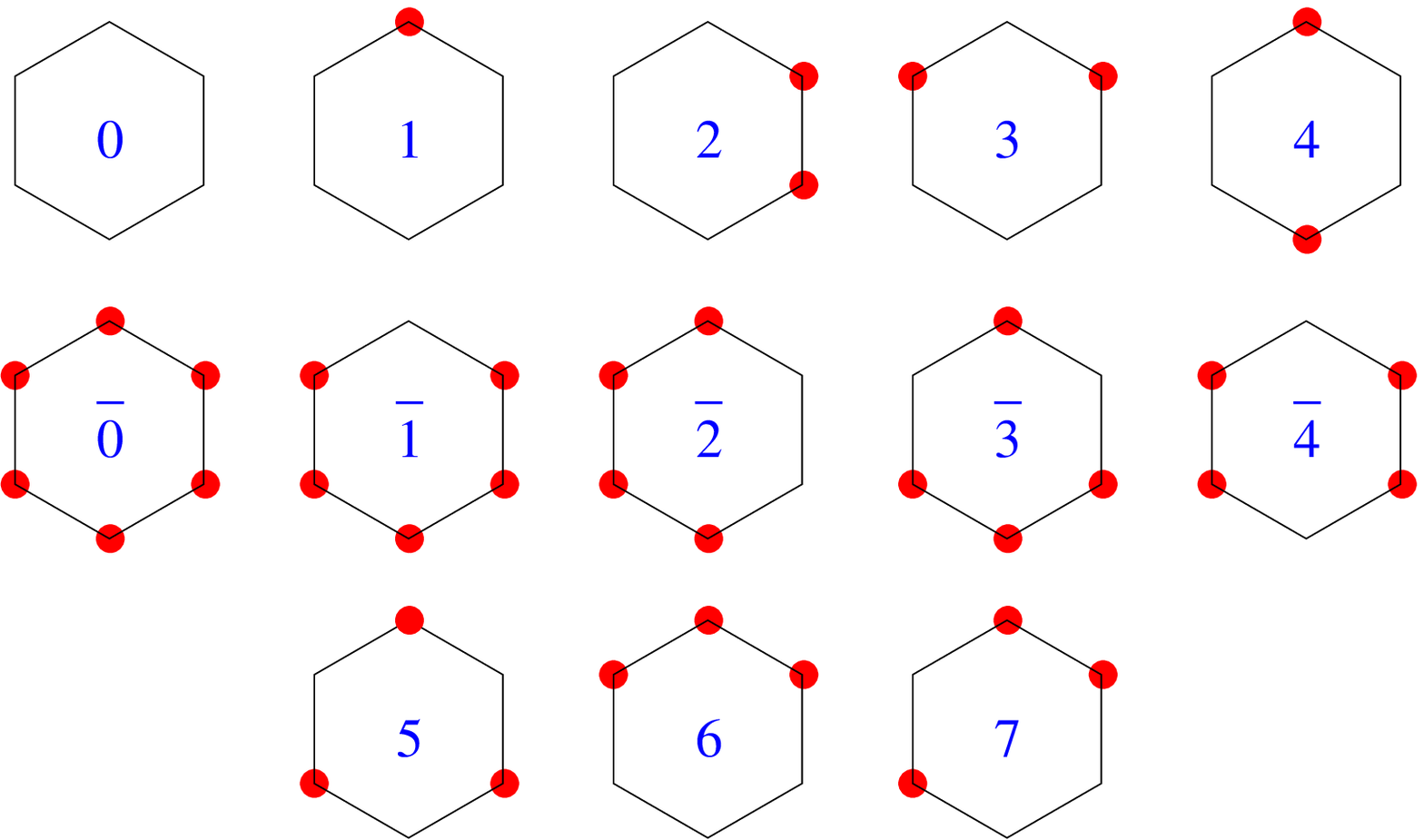}}
	\subfigure[square plaquette]{
	\label{fig:plaquette_types3}		
        \includegraphics[width=2.0in]{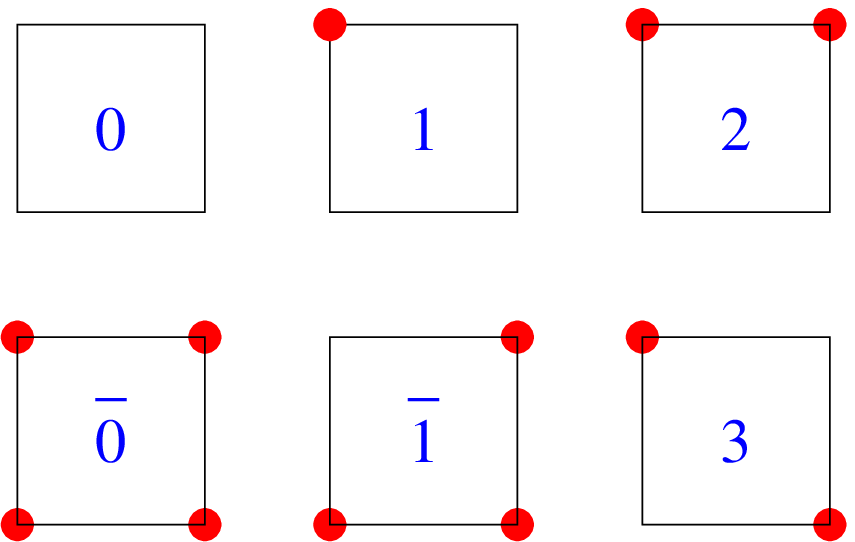}}
	\label{fig:plaquettes}
	\caption{(Color online) Hexagonal and square plaquette Ising configurations. The minority sites are denoted by solid (red) circles.}
\end{figure}

Remarkably, the manipulations in the previous section imply that, as a
function ${\mathcal E}_{\mathcal P}(\sigma_1, \cdots, \sigma_L)$ is
the \emph{same} for {\sl all} lattices with the same plaquette size
(4,6,8 etc.), for any $k$.  The differences arise in the allowed
plaquette configurations, which depend strongly on $q$ and $k$.
This function is most conveniently described by giving the energies for
all possible plaquette configurations. For the lattices with hexagonal
plaquettes we find 
\begin{eqnarray}\label{hexagonal_ergs} \hspace{-1.0in}
{\mathcal E}_0 = & {\mathcal E}_{\bar 0} = 0, \hspace{2.0in}
{\mathcal E}_1 = {\mathcal E}_{\bar 1} = \frac{1}{4}{\mathcal E}_{2} = \frac{1}{4}{\mathcal
  E}_{\bar 2} = -\frac{J s^3 \alpha
  ^6}{32 (1 - 4 s)^2} ,
\\ \hspace{-1.0in}
{\mathcal E}_3 = & {\mathcal E}_{\bar 3} = -\frac{J s^4 \left(512 s^2 -
    256 s + 33\right) \alpha ^6}{16 (3 - 8 s)^2 (4 s - 1)^3}, \hspace{0.8in}
{\mathcal E}_4 = {\mathcal E}_{\bar 4} = -\frac{J (1 - 8 s)^2 s^4 \alpha ^6}{8 (4 s - 1)^5}, \nonumber
\\ \hspace{-1.0in}
{\mathcal E}_5 = & -\frac{6 J s^5 \left(3072 s^5 - 4352 s^4 + 2432 s^3 -
    660 s^2 + 88 s - 5\right) \alpha ^6}{(3 - 8 s)^2 (4 s - 1)^5
  \left(24 s^2 - 22 s + 5\right)}, \hspace{0.5in}
{\mathcal E}_6 = 6 {\mathcal E}_{\bar 1}, \nonumber
\\ \hspace{-1.0in}
{\mathcal E}_7 = & -\frac{J s^4 \left(128 s^2-40 s+3\right) \alpha ^6}{8 (1-4 s)^4 (8
   s-3)} -   
   \frac{4 J s^6 \left(5184 s^3-3600 s^2+789 s-56\right)
   \alpha ^6}{3 (4 s-1)^5 \left(64 s^2-32 s+3\right)^2 \left(72
   s^2-18 s+1\right)} \nonumber
   \\ \hspace{-1.0in}
   - &  
   \frac{J s^4 \alpha ^6}{2 (1-4 s)^2 (8 s-3)}
\; . \nonumber
\end{eqnarray}
The various plaquettes types are shown in figure~\ref{fig:plaquette_types2}.
For the pyrochlore $k=2$ plateau, all 13 plaquette configurations in
figure~\ref{fig:plaquette_types2} are possible.  Not surprisingly, the
energies for plaquette configurations with up and down spins swapped
are identical, since in this case ($k=2, q=4$) there are equal numbers
of minority and majority sites on each simplex.  Interestingly, this
case can be interpreted as an ``ice model'', in which the two down spins
on  a tetrahedron indicate the locations of protons on the oxygen atom
at its center.

For the kagome $k=1$ antiferromagnet at $1/3$ magnetization and the
pyrochlore $k=1$ antiferromagnet at $1/2$ magnetization (the subject
of \cite{Bergman4}), there are only 5 allowed plaquette
configurations, those with no two neighboring minority sites. These
are the plaquettes configurations labeled $0,1,3,4,5$. Their energies
are the same expressions as in \ref{hexagonal_ergs}.

For lattices with square plaquettes, we find
\begin{eqnarray}\label{square_ergs} \hspace{-1.0in}
{\mathcal E}_0 = & {\mathcal E}_{\bar 0} = 0, 
\hspace{1.9in}
{\mathcal E}_1 = {\mathcal E}_{\bar 1} = -\frac{J s^3 \alpha ^4}{2 (4 s - 1)^2}, 
\\ \hspace{-1.0in}
{\mathcal E}_2 = & -\frac{J s^3 \left(64 s^2 - 24 s + 3\right) \alpha^4}{2 (4 s - 1)^3 (8 s - 1)}, 
\hspace{0.5in}
{\mathcal E}_3 = -\frac{8 J s^4 \left(8 s^2 - 5 s + 1\right) \alpha ^4}{(4 s - 1)^3 \left(16 s^2 - 14 s + 3\right)} 
\; . \nonumber
\end{eqnarray}
The square plaquettes types are described in
figure~\ref{fig:plaquette_types3}. The $k=2$ (zero magnetization) plateau of the checkerboard lattice
allows all 6 plaquette types outlined in
figure~\ref{fig:plaquette_types3}. For $k=1$ ($1/2$ magnetization),
only types $0,1,3$ are allowed.

As explained in detail in \cite{Bergman4}, these energies
are actually slightly redundant.  To see this, first define $x_a$ as
the fraction of plaquettes in the lattice in configuration $a$.  Then the energy per plaquette of any
state is given by 
\begin{equation}
  \label{eq:Hdx}
  {\mathcal H}_{\textrm{diag}}/N = \sum_a x_a {\mathcal E}_a,
\end{equation}
where $N$ is the total number of plaquettes.  Now one notices that the
fractions obey two global constraints: first the fractions must sum to
$\sum_a x_a =1$, and second, the total fraction of minority sites must
be $\frac{k}{q}$.  Denoting the fraction of minority sites in each
plaquette configuration by $M_a$, the latter constraint implies
$\sum_a x_a M_a = \frac{k}{q}$.  Because of these two constraints, we
see that it is possible to shift the energies in equation~\ref{eq:Hdx} in
such a way as to change the energy per plaquette by a constant for all
allowed configurations: ${\mathcal E}_a \rightarrow c_0 + c_1
M_a$, with $c_0$ and $c_1$ {\sl arbitrary}.  By appropriate choice of
these constants, one can tune two plaquette configuration energies to
zero.

\section{Lowest energy states for the $k=1$ effective Hamiltonians}
\label{sec:lowest-energy-states}

The plateau states described above can always be mapped to (sometimes
overlapping) dimer coverings on lattices with sites at the center of
each simplex, and links passing through every site.  Each site in the
dimer lattice therefore has $q$ links emanating from it. A minority site
in the original lattice corresponds to the presence of a dimer on the
corresponding link in the dimer lattice.  With $k$ minority sites, each
site in the dimer lattice is covered by $k$ dimers.  The diagonal term
in each effective Hamiltonian simply assigns different energies to the
various dimer coverings.  For the cases with $k=1$, which correspond to
plateau states with non-vanishing magnetization, these models are of the
form of generalized Quantum Dimer Models (QDMS)\cite{RK:prl88}, though
often with a modified and more complicated diagonal term.  In this
section, we use known results for these QDMs and simple optimization
arguments to analyze the particular examples for which we have derived
the effective Hamiltonian in the previous section.

\subsection{Checkerboard lattice}

First we address the simplest effective Hamiltonian -- the one obtained
for the checkerboard lattice.  The dimer lattice is the square
lattice, and so for $k=1$ the Hilbert space of dimer coverings is the
same one of \cite{RK:prl88}.  

For $k=1$, there are only three plaquette configurations possible, so
we shift energies so that only the energy of the flippable plaquette
(type 3) is non-zero and equal to $V =  {\mathcal E}_0 + {\mathcal E}_3 - 2 {\mathcal E}_1$.
For $s > \frac{1}{2}$ it is straightforward to show that $V<0$. For
$s=\frac{1}{2}$ there is a divergence, which indicates that the
procedure is invalid in that case.  For spin-$1/2$, this occurs
because the off-diagonal term appears already at
order $\alpha^2$, whereas the diagonal term appears at order
$\alpha^4$.  Because off-diagonal processes are possible already at
second order, some of the intermediate projection operators in
equation~\ref{eq:effshrod} must be treated more carefully in that case.
However, for $s=\frac{1}{2}$ the
off-diagonal term is in any case dominant, so the diagonal term can be
neglected.  Therefore we take $V=0$ in this case, and for all $s \geq
\frac{1}{2}$, we have $V \leq 0$.

Adding the lowest order off-diagonal term, which flips between the two
type 3 plaquette configurations on a given plaquette, we find the
lowest order diagonal and off-diagonal terms form a QDM of
\emph{exactly} the same form of \cite{RK:prl88}
\be\label{Square_QDM}
{\mathcal H}_{\textrm{QDM}}  =
V \sum_{\mathcal P}
\left(
{\centering \includegraphics[width=1.4in]{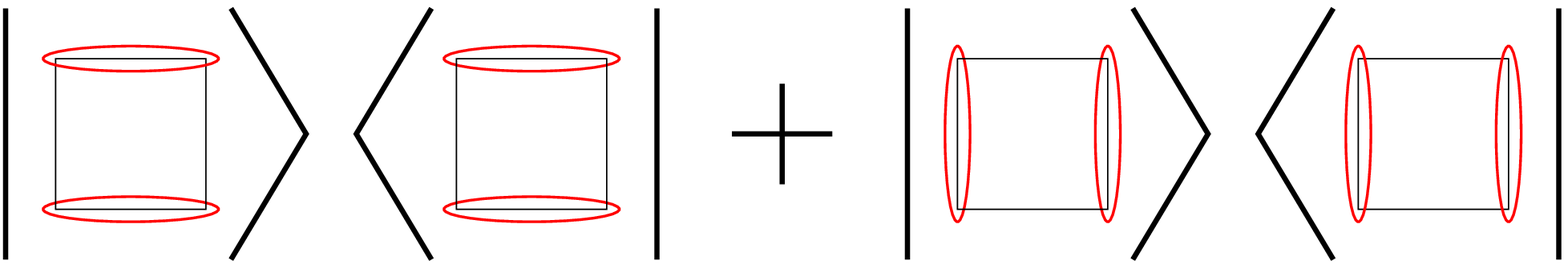}} 
\right)
- t \sum_{\mathcal P}
\left(
{\centering \includegraphics[width=1.4in]{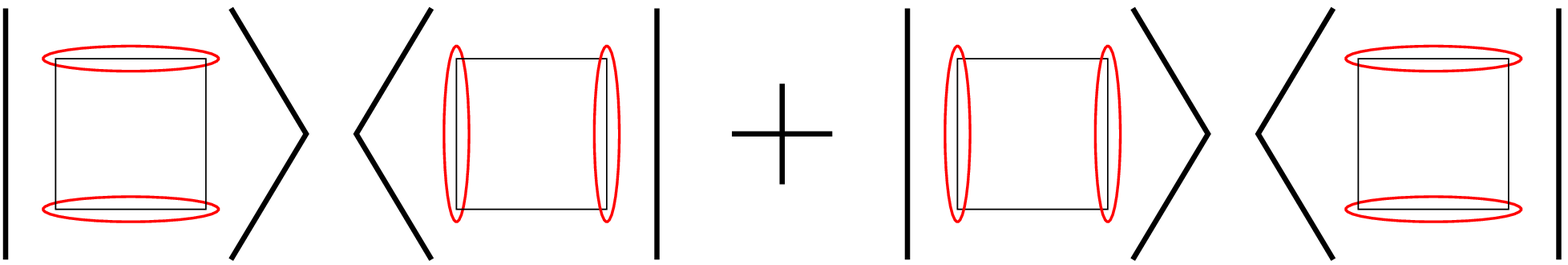}} 
\right)
\; ,
\ee
where $t = K_4(s) \alpha^{4 s} J > 0$.

The model \ref{Square_QDM} has undergone in-depth numerical
scrutiny (see \cite{Syljuasen}), which shows that for $V/t < +0.6 \pm
0.05$ one obtains a columnar phase, in which the dimers preferentially
sit on staggered columns of parallel bonds. For any value of $s$, we
have $V/t \leq 0$.  Therefore, the ground state is always in the
columnar phase.
\subsubsection{Kagome lattice}

Next we turn to the kagome $k=1$ plateau, in which the magnetization
is $\frac{1}{3}$ of the full polarization possible. The dimer lattice
in this case is the honeycomb lattice, and so for $k=1$ the Hilbert
space of dimer coverings is that of the QDM of
\cite{Moessner:prb01}.  The effective Hamiltonian takes on
the form
\be\label{honeycomb_QDM}
{\mathcal H}_{\textrm{QDM}} =
\sum_{\mathcal P} {\mathcal E}_{\mathcal P}
- t \sum_{\mathcal P}
\left(
{\centering \includegraphics[width=1.4in]{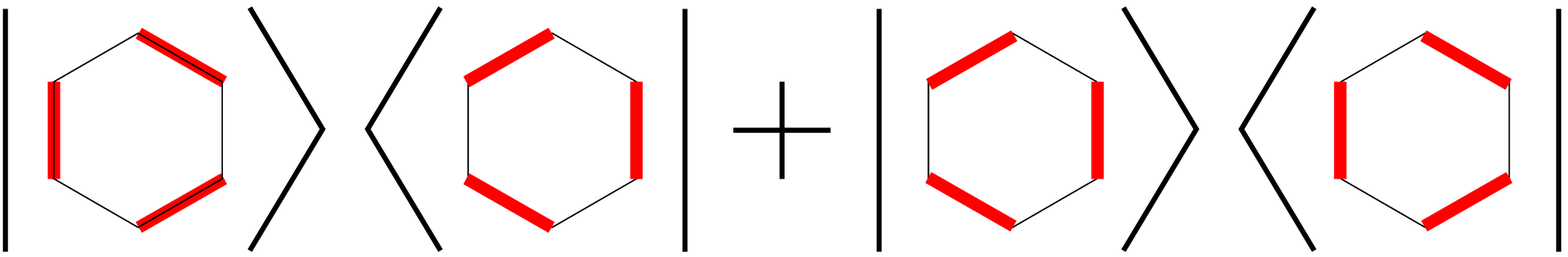}} 
\right)
\; ,
\ee
with the plaquette energies defined in
equation~\ref{hexagonal_ergs}, and $t = K_6(s) \alpha^{6 s} J > 0$.
As before, for spin $s=\frac{1}{2}$ the diagonal energies are taken to
be $0$.  In that case, the model reduces to a particular case of the
model considered in \cite{Moessner:prb01}, in which they
find the phase is a ``plaquette valence bond solid'', in which
resonating plaquettes (superpositions of the two type 5 configurations)
form a $\sqrt{3}\times\sqrt{3}$ sublattice of all plaqettes.

We can tune the honeycomb lattice plaquette energies so that types
$0$ and $3$ have energy $0$. The remaining plaquette energies are 
$V_5 = \frac{1}{2} \left(\mathcal{E}_0+2
  \mathcal{E}_5-3\mathcal{E}_3\right)$, $V_1 =  \frac{1}{2}
\left(-\mathcal{E}_0+2\mathcal{E}_1-\mathcal{E}_3\right)$, $V_4 =
\mathcal{E}_4-\mathcal{E}_3$. $ $From equation~\ref{hexagonal_ergs}, we find that for $s \geq 1$, $V_5 <0$ and
$V_{1,4} >0$, so that the flippable plaquettes (type 5 in
figure~\ref{fig:plaquette_types2}) are favored energetically.

Ignoring the off-diagonal term for a moment, the lowest energy dimer
covering configuration for the entire range of $s$ has the $\sqrt{3}
\times \sqrt{3}$ structure of the columnar state of
\cite{Moessner:prb01}.  This dimer covering includes
only type $0$ and type $5$ plaquettes, with $\frac{1}{3}$ of the
plaquettes in the type $0$ configuration, and the remaining
$\frac{2}{3}$ plaquettes in the type $5$ configuration.  From the
constraint on the total fraction of the minority sites $\sum_a x_a M_a
= \frac{1}{3}$ we can immediately deduce that $x_5 M_5 \leq
\frac{1}{3}$ since all $x_a \geq 0$ and all $M_a \geq 0$.  With $M_5 =
\frac{1}{2}$ we find that $x_5 \leq \frac{2}{3}$, so that $\sqrt{3}
\times \sqrt{3}$ state has the maximum fraction of type $5$
(flippable) plaquettes we can pack on the lattice, already an
indication that this is a very good energy state. Since the remaining
plaquettes are of type $0$ with energy $0$, which is the second best
energy possible for any plaquette configuration, this is clearly the
lowest energy possible for any dimer covering in this model.

Next we will try to analyze the model \ref{honeycomb_QDM} with the
off-diagonal included.  Since the diagonal term favors flippable
plaquettes, it is not unreasonable to approximate $V_{1,4} \approx 0$.
Then our model becomes the same as the QDM of
\cite{Moessner:prb01}
\be\label{pure_QDM}
{\mathcal H}_{\textrm{QDM}}  =
V \sum_{\mathcal P} 
\left(
{\centering \includegraphics[width=1.4in]{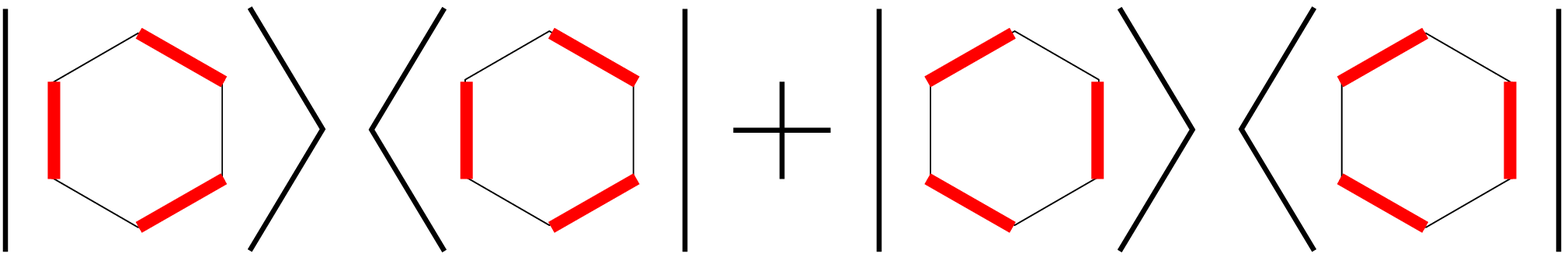}} 
\right)
- t \sum_{\mathcal P}
\left(
{\centering \includegraphics[width=1.4in]{honeycomb_QDM.eps}} 
\right)
\; ,
\ee
where $V = V_5 \leq 0$.  Calculating the
off-diagonal coefficient $t$ by the methods of section~\ref{sec:Model},
we obtain $\alpha^{6(s-1)}V/t  =
-0.246927,-2.24511,-45.155,-1228.23$ respectively for $s=1,3/2,2,5/2$.
For all $s > 1$, even when extrapolating $\alpha \rightarrow 1$, we
find that the ratio $V/t$ takes on a value that puts it in a
columnar valence bond solid phase, according to the results of
\cite{Moessner:prb01}.  For $s = 1$, the diagonal and
off-diagonal terms are of the same order in $\alpha$, and the ratio
$V/t\approx -0.247$ is $\alpha$-independent.  However, this 
puts the system described by equation~\ref{pure_QDM} within the quoted error bar
for the transition point between the plaquette and columnar valence
bond solid phases in \cite{Moessner:prb01}: $V/t = -0.2 \pm
0.05$.  The $s = 1$ case may therefore be more sensitive to the values
of $V_{2,4}$ than other values of $s$.

\subsection{Pyrochlore lattice}

Finally we turn to the case of the pyrochlore lattice.  The $k=1$ case
was discussed in \cite{Bergman4}, but we will give a brief
summary here.  As discussed in section~\ref{sec:results}, the plaquette
energies are off-diagonal term are {\sl identical} to those on the
simpler kagome lattice.  Thus the effective Hamiltonian is the same as
the one given in equation~\ref{honeycomb_QDM}, but with dimers on links
of the diamond lattice.  For sufficiently large $s$, the off-diagonal
term can be neglected, and one need only find the classical dimer
covering minimizing the diagonal energy.  For large $s\gg 1$, this
classical optimization problem could be analytically solved.  The
resulting ``trigonal$_7$'' state has trigonal symmetry and a seven-fold
enlarged magnetic unit cell.  Numerical investigations estimate that
this state obtains for $s\geq 2$.  

For spin $s=3/2$, the situation is more complex.  Ignoring first the
off-diagonal term, the diagonal term alone does not fully split the
degeneracy of the dimer manifold, and the ground state subspace grows
exponentially with the system length.  If an extrapolation is made to
the isotropic limit $\alpha=1$, the off-diagonal term is
non-negligible.  One may approximately include its effects by using
equation~\ref{pure_QDM}, as discussed above for the kagome lattice.  In
this model, two likely ground states were proposed in
references\cite{Bergman4},\cite{Bergman:prl05}: the ``{\bf R}'' state,
with a 4-fold enlarged simple cubic magnetic unit cell, and a
``resonating plaquette state'', with more complex structure.  

For $s=1$ and $s=1/2$, the off-diagonal term is dominant.  While the
above {\bf R} and resonating plaquette states remain candidates, another
possibility is a $U(1)$ spin liquid state, which is known to be the
ground state of equation~\ref{pure_QDM} for $V/t$ close to but less than
$1$.  

\ack

The authors acknowledge discussions with Kedar Damle.  This work was
supported by NSF Grant DMR04-57440, PHY99-07949, and the Packard
Foundation. R.S. is supported by JSPS as a Postdoctoral Fellow.

\end{document}